\documentclass{tMOP2e} 

\usepackage{epsfig}

\begin{document}

\title{Continuous-wave phase-sensitive parametric image amplification}

\author{S. GIGAN, L. LOPEZ, V. DELAUBERT, N. TREPS, C. FABRE and A. MAITRE}

\thanks{Laboratoire Kastler-Brossel, Université Pierre et Marie Curie, Case 74, 75252 PARIS cedex 05. gigan@spectro.jussieu.fr}

\maketitle

\begin{abstract}

We study experimentally parametric amplification in the continuous
regime using a transverse-degenerate type-II Optical Parametric
Oscillator operated below threshold. We demonstrate that this
device is able to amplify either in the phase insensitive or phase
sensitive way first a single mode beam, then a multimode image.
Furthermore the total intensities of the amplified image projected
on the signal and idler polarizations are shown to be correlated
at the quantum level.

\end{abstract}

\section{Introduction}

Parametric interaction is widely used in quantum optics, because
it is a very useful tool to produce quantum correlated and
entangled light. In most cases, only temporal or polarization
aspects of entanglement or squeezing are used\cite{bachor}. But
the parametric process also produces spatial quantum entanglement
\cite{Jedrkiewisz} which can be useful in
many applications, for example in two-photon or "ghost" imaging
\cite{Gatti:04}, a subject which has recently aroused a great
interest in the community.

Due to the low parametric coupling in crystals, experiments with
parametric crystals are usually operated in the photon-counting
regime, or require the high peak power of pulsed lasers. However
many applications, such as high sensitivity measurements below
shot noise, require intense non-classical c.w. beams: instead of
single pass parametric interaction, one then prefers Optical
Parametric Oscillators which produce quantum-correlated laser-like
beams\cite{heidmann}. Usually the cavity used for the OPO is resonant
for a single transverse mode of the field and acts as a perfect
spatial filter. In order to obtain imaging effects in OPOs, one
needs cavities which are simultaneously resonant on many
transverse modes, the so-called transverse-degenerate cavities,
such as the concentric, confocal or hemi-confocal cavities.
Classical multimode effects, such as pattern formation, have been
observed in transverse degenerate OPOs \cite{Ducci:01}.

Quantum spatial effects in OPOs have been the subject of many
theoretical papers \cite{Lugiato1,Petsas}. As far as experiments are concerned,
the first multimode non-classical effect observed in OPOs has been
reported in \cite{Martinelli}. By measuring the transverse
distribution of correlations between the signal and idler beams
emitted by a confocal OPO above threshold, it was experimentally
shown that this device generated a non-classical multimode beam.

The present paper reports experimental results obtained in a
frequency degenerate OPO operating below the oscillation
threshold. The aim of the experiment is to use a
transverse-degenerate OPO to amplify a weak injected image in a
phase-sensitive way, in order to realize noiseless amplification
and quantum cloning of the original image. This work is an
extension to the c.w. regime to what has been already observed in
the pulsed, single pass regime \cite{Kumar:99,Lantz}. It presents
the first experimental achievements obtained with this device:
observation of c.w. single-mode and multimode parametric
amplification at the classical and quantum level.

The paper is organized as follows: after a short reminder of the
main features of type II parametric amplification, section III
reports on the observation of single-mode amplification of a weak
injected signal in the phase sensitive configuration, first
using a scanned cavity, then in a locked cavity configuration.
It then describes a new set-up based on a dual-cavity, where pump
and signal-idler resonate in different cavities, and show that
stable phase-sensitive amplification is obtained in this device.
In section IV, we demonstrate multimode amplification of an image
made of a double slit with the same system operating in a
hemi-confocal configuration, and show finally the existence of
quantum correlations between the total intensities of the signal
and idler amplified images.

\section{Properties of type II, frequency degenerate, Optical
parametric Amplification}

Let us consider a type II parametric interaction where the signal
and idler modes have the same frequency $\omega$ and are
respectively polarized along the $Ox$ and $Oy$ axes. A weak field
of the same frequency $\omega$ injected in such a device is
predicted to be amplified at the output of the crystal. If the
input field polarization is aligned either along the $Ox$ or $Oy$
axis, the amplification is phase insensitive and therefore
inevitably adds noise to the input field. The signal to noise
ratio is deteriorated by a factor equal to 2 in the high gain
limit. In contrast, when the input field is injected at 45 degrees
from these axes, the amplification is phase sensitive: there is
either amplification or de-amplification, depending on the
relative phase between the pump wave at $2\omega$ and the injected
wave \cite{Levenson}. In the ideal case, there is no
noise added in the latter configuration: the amplification is said
to be noiseless. Furthermore, if one decomposes the output field
on its two polarization components along the signal and idler
polarization axes $Ox$ and $Oy$, one gets two intensity correlated
outputs, which are in some respect two "quantum clones"
\cite{Grangier}. These effects have already been experimentally
demonstrated \cite{Levenson} for a single transverse mode input
field  in single pass amplification using an intense pulsed laser
as a pump. To extend it to the domain of continuous wave fields,
one puts the crystal in a resonant cavity and go to the
"regenerative parametric amplification" regime, with a gain
tending to infinity when one approaches the oscillation threshold
of the Optical Parametric Oscillator from below. The modulation
bandwidth of this amplifier is reduced with respect to the single
pass amplifier, and restricted to the optical cavity bandwidth. At
the quantum level \cite{Protsenko}, the intracavity parametric
amplifier has essentially the same properties as the single pass
one: possibility of noiseless amplification and quantum cloning.
Finally, if one wants to reduce the threshold, in order to be able
to pump the system high above threshold with moderate powers and
excite at the same time many transverse modes, it is better to use
a triply resonant OPO, where the signal, idler and pump modes are
simultaneously resonating.

\section{Single mode c.w. optical amplification}

\subsection{Experimental Setup}
\begin{figure}[htbp]
\begin{center}
\epsfig{file=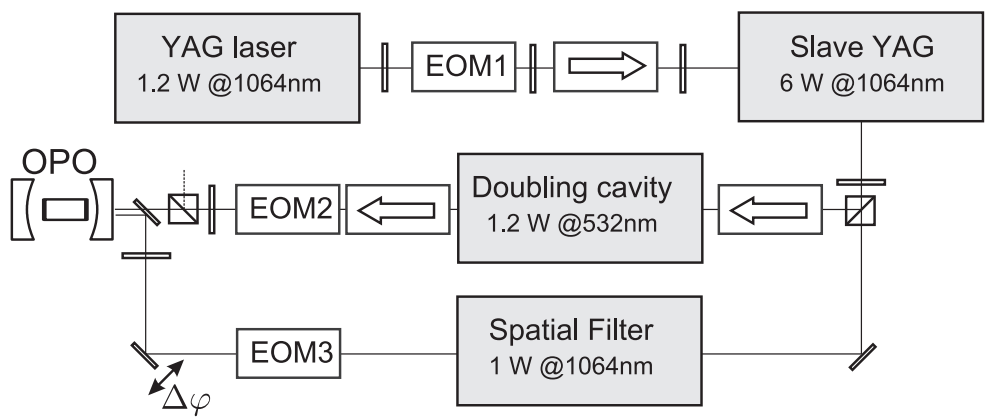,width=1\linewidth} \caption{Source system
for amplification.} \label{sources}
\end{center}
\end{figure}

The experimental setup is an upgrade of the one described in
\cite{Martinelli} (figure \ref{sources}). We start from a very
stable Nd-YAG master laser from InnoLight, delivering 1.2W
single-mode continuous output power at 1064 nm, which is used to
inject and stabilize a flash-lamp-pumped ring-cavity Nd-YAG
"slave" laser. This configuration gives a stable single-mode beam
of up to 6W at 1064 nm. A part of it is injected into a
semi-monolithic frequency doubling cavity, using a MgO:LiNbO$_3$
crystal, which produces up to 1.7 W of $532nm$ green light, used
as a pump beam for the OPO. The remaining part is injected into an
impedance-matched ring Fabry-Perot cavity used as a spatial filter
("mode cleaner"), and used to inject the amplifier. An object,
such as a slide or a resolution target, can be placed in the beam.
With this system it is possible to obtain 1.2W at 532 nm for pump,
and simultaneously and 1W at 1064 nm for injection. The phase
between the pump and the injection is controlled by mirror placed
on a piezo-electric transducer.

The different optical cavities are stabilized using the
Pound-Drever-Hall method \cite{Black:01}. For this purpose three
electro-optic phase modulators (EOM) are inserted in the beam
path. EOM1 adds a phase modulation at $14.7MHz$ on the infrared
beam and allows to lock the slave laser, the doubling cavity and
the spatial filtering cavity. EOM2 adds a phase modulation at 19
MHz on the pump beam to lock the OPO on a pump resonance. EOM3
modulates at 16.5 MHz the infrared injection beam to stabilize the
OPO on a signal or idler resonance.

\begin{figure}[htbp]
\begin{center}
\epsfig{file=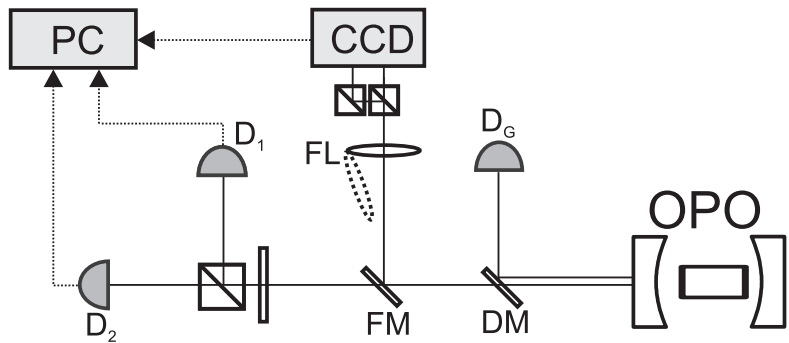,width=.8\linewidth} \caption{detection
scheme.} \label{detection}
\end{center}
\end{figure}

The detection scheme  is represented on figure \ref{detection}: a
dichroic mirror (DM) rejects the residual green light from signal
and idler modes. The reflected pump is sent onto a photodiode
($D_G$) to monitor the OPO and to stabilize it. Signal and idler
beams are separated by a polarizing beam splitter and sent to two
InGaAs photodetectors ($D_1$ and $D_2$), having matched quantum
efficiencies within better than 1$\%$ and equal to $90\%\pm5\%$. A flipping mirror (FM) with
approximately 50 $\%$ reflection diverts some of the intensity of
the signal and idler beams onto an imaging set-up, which allows us
to simultaneously record on a CCD camera the signal and idler
transverse intensity distributions. A secondary flipping lens (FL)
is used to switch between two configurations: in the first one,
the CCD plane is the conjugate plane of the crystal center plane  ("near field") ; in the
second one, one monitors on the CCD the Fourier transform of the
previous one ("far field").

The data acquisition system relies on computer storage of the
quantum fluctuations: the photocurrent, amplified by a low-noise
36dB broadband transimpedance amplifier, is mixed with a current
oscillating at frequency $5MHz$ and filtered on a $100kHz$
low-pass filter. Together with the DC photocurrents, the high
frequency signals are then recorded by a acquisition card
(PCI6110E from National Instrument), which stores in four 12-bit
channel at a rate of 200 kHz the DC and the demodulated HF parts
of each photocurrent. A post-processing treatment of the data
allow us first to calibrate the shot noise level and then to
obtain normalized noise variances on the signal and idler
photocurrents normalized to shot noise, together with their
normalized correlation function.

\subsection{Single mode amplification in a nearly confocal cavity}

We use a walk-off compensated KTP crystal cut for non-critical
type-II colinear phase-matching and frequency degeneracy. Its
temperature is electronically stabilized around 33 $^\circ C$
within 1 mK. The cavity is made by two mirrors with radii of
curvature $R=100$ mm, separated by a distance of approximately 100
mm, i.e. close to confocality, but not exactly tuned to it. The
input coupler reflects 90 $\%$ of the 532 nm light, and almost all
the 1064 nm light. The output coupler reflects almost all the 532
nm light, and $99\%$ of the 1064 nm light. The oscillation
threshold of this OPO is approximately 20 mW. The operating point
is taken as close as possible to the threshold for higher
parametric gain.

Both pump and injection are spatially matched to the $TEM_{00}$
cavity mode, and a half-wave plate allows us to rotate the
polarization of the injection beam, and to inject it on a chosen
polarization direction. As the refractive indices for signal,
idler and pump (respectively 1.8296, 1.7467 and 1.7881) are
different, the three modes do not in general resonate for the same
length of the cavity. Triple resonance is achieved by precisely
tuning various cavity parameters. A fine temperature tuning gives
simultaneous resonance of signal and idler modes. The temperature
range to get it is approximately 20 mK, within our regulation
stability range. Tilting the crystal allow us to reach the triple
resonance configuration. Only a very slight tilt is possible
without deteriorating the optical alignment of the cavity.

This method has allowed us to observe a strong amplification in
the phase insensitive configuration (injection along the signal or
idler polarisation), while sweeping the cavity length through the
triple resonance point. We have measured a maximum gain in such a
"transient" c.w. Optical Parametric Amplifier of $23dB$.

\begin{figure}[htbp]
\begin{center}
\epsfig{file=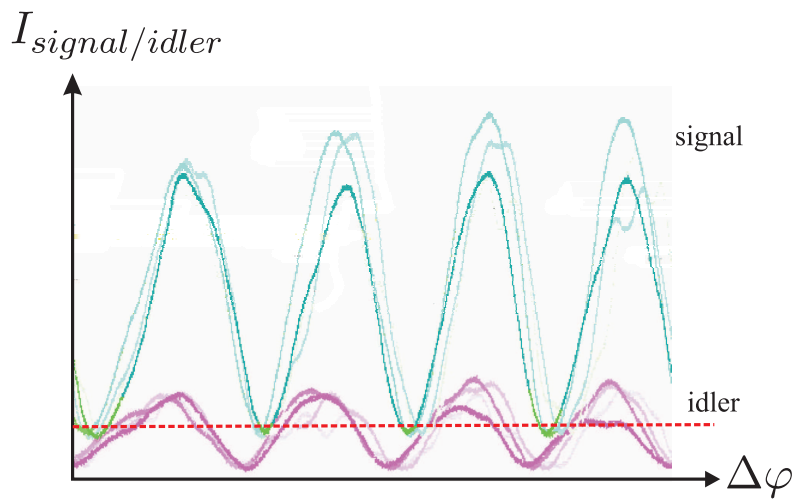,width=.8\linewidth}
\caption{Signal and idler intensities as a function of the
relative phase between the injected $TEM_{00}$ mode and the pump
mode. The input field polarization is at 45 degrees from the
cavity optical axes. The dashed line is the signal and idler
intensity without amplification.} \label{PSAlockeconfocal}
\end{center}
\end{figure}

We have then rotated the injected field polarization by 45 degrees
and stabilized the OPO length on the triple resonant point. By
scanning the relative phase between the pump and the injected
field, we have observed phase sensitive amplification with gain up
to 6dB (figure \ref{PSAlockeconfocal}) on time scales of the order
of a few seconds. To our knowledge, this is the first observation
of such an effect in the c.w. regime and in a type II
configuration. The gain value is smaller than in the scanned
configuration, because of crystal heating effects which appear
when the cavity is locked and bring the signal and idler modes out
of resonance, and also because exact and stable triple resonance
is difficult to obtain. As can be seen on figure
\ref{PSAlockeconfocal}, the signal and idler beams do not have the
same mean intensity because the idler mode is not perfectly
resonant. This problem of triple resonance is very similar to the problem of frequency degenerate operation of an OPO above threshold  \cite{Laurat,Pfister}. 

\subsection{Single mode amplification in a dual cavity}

To solve the stability problems we have used a new setup, based on
a dual cavity configuration, in which the signal and idler modes
resonate in a different cavity from the pump cavity, but where the
three modes overlap in the crystal. To our knowledge this kind of
cavity has been introduced and realized for the first time in
\cite{Colville:94}, and has been used in a c.w. OPO in
\cite{Wong:98} to produce twin beams, in the pulsed
regime\cite{Rosencher:00} to emit single mode beams or to produce
mode-hop free continuously tunable OPO\cite{Turnbull:00}.

\begin{figure}[htbp]
\begin{center}
\epsfig{file=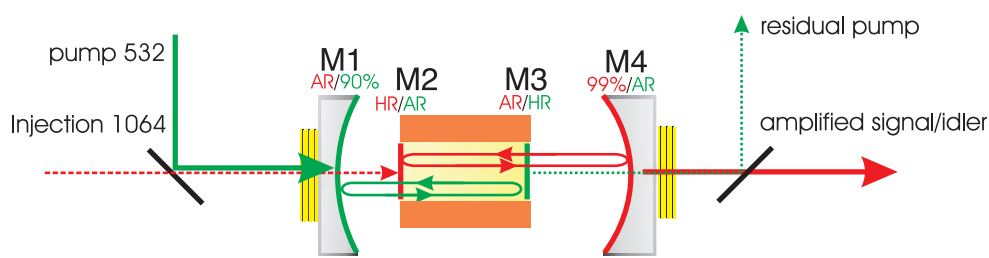,width=1\linewidth}
\caption{Scheme of the
dual-cavity.} \label{SchemaPrincipeDC}
\end{center}
\end{figure}

The linear dual cavity is shown in figure \ref{SchemaPrincipeDC}.
The pump cavity is limited by mirrors M1 and M3, and the
signal-idler cavity by mirrors M2 and M4. The characteristics of
these mirrors are summarized in table \ref{TraitementsMiroirsDC}.
Both cavities are approximately 50mm long, and the 10mm-long KTP
crystal is cut for frequency degenerate operation, and
non-critical phase-matching, but is not walk-off compensated. The
advantage of such a dual cavity is that it is now possible to
control and stabilize independently the length of each cavity. The
triple resonance condition is then obtained by temperature tuning,
and is much more stable. M2 and M3 being parallel, the parallelism
of the optical axes of the two cavities is automatically ensured.
The coincidence of these axes is finally adjusted, by slightly
tilting the mirrors M1 or M4 to maximize the mode overlap, so as
to obtain the minimum threshold possible, of the order of $30mW$.

\begin{table}[htbp]
\begin{center}
\begin{tabular}{l|c|c|c|c}
    & M1  & M2 & M3 & M4 \\ \hline
    Radius of curvature  & $100 mm$ & plane &  plane & $100 mm$  \\
  reflectivity at $532 nm$    & 90\% & 5.25 \% & 99.3 \% & 6.6\% \\
  reflectivity at $1064 nm$  & 5\% & 99.96 \% &  0.11 \% & 98.93 \%\\
\end{tabular}
 \caption{characteristics of the mirrors of the dual cavity}\label{TraitementsMiroirsDC}
\end{center}
\end{table}

Because of the imperfect antireflection coatings deposited on the
crystal faces, unavoidable with dual-wavelength coatings, the
M1-M2 and M3-M4 systems constitute low finesse secondary cavities.
This implies in particular that the intracavity pump signal and
idler powers depend also on the resonance condition of these
cavities, so that it is difficult to know the exact power of the
modes circulating in the cavity.

\begin{figure}[htbp]
\begin{center}
\epsfig{file=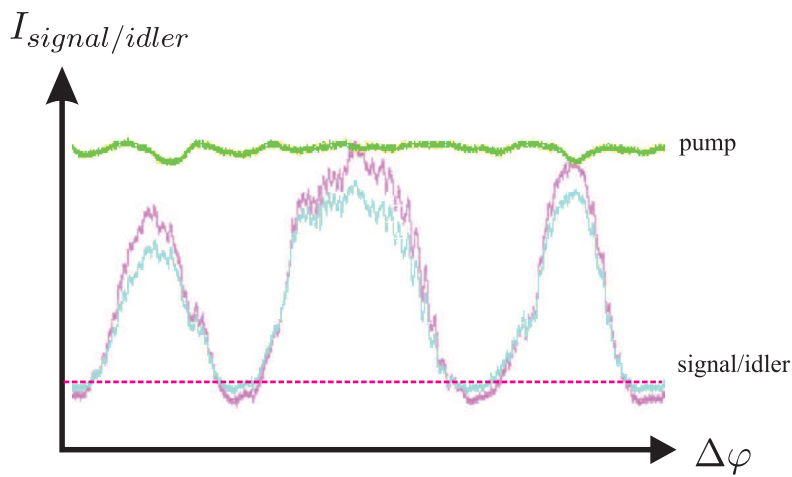,width=.8\linewidth} \caption{Signal and
idler intensities as a function of the relative phase between the
injected $TEM_{00}$ mode and the pump mode. The input field
polarization is at 45 degrees from the cavity optical axes. The
geometry of the dual cavity is close to hemi-confocal. The dashed
line is the signal and idler intensity without amplification.}
\label{ampliDC1}
\end{center}
\end{figure}

When the dual cavity is injected at 45 $^\circ$ of the crystal
axes, and when the pump and injected modes are the $TEM_{00}$
modes of their respective cavities, we have measured below the
oscillation threshold c.w. stable phase-sensitive amplification
with typical gains of 5-6 dB (figure \ref{ampliDC1}). The precise
gain value is not known because of the secondary cavity effect.
Let us note that in this new set-up, the signal and idler modes
have the same intensity, and that its remains stably on the triple
resonant point during several minutes. It was not possible to
observe higher gains because the intracavity pump intensity
variations due to the secondary cavity prevented us from operating
very close to threshold.

When one removes mirror M1, one obtains a doubly resonant OPO,
having an oscillation threshold around $200 mW$. We have also
observed phase sensitive amplification with typical gain of 4-6dB in
this configuration. We have evaluated finally the quantum
correlation between the intensities of the signal and idler
components of the amplified output beam : for a parametric gain of
the order of 6dB, the noise on the difference between the two
intensities has been measured 35 $\%$ below the standard quantum
limit. These results are consistent with the expected performance
of the cavity and of the detection scheme, taking into account
losses and quantum efficiency.

\section{multimode amplification}

\subsection{Experimental setup}

The previous set-up has been modified for multimode image
amplification: by using a mirror M1 having a large radius of
curvature, the pump cavity $TEM_{00}$ mode has a larger waist, and
is likely to pump many transverse modes of the signal-idler
cavity. This latter cavity is then brought to an exact transverse
degenerate configuration. Because of the plane surfaces of the
mirrors M2 and M3 deposited on the crystal faces, we have been
compelled to  use the hemi-confocal configuration, in which the
concave mirror is separated from the plane mirror by a distance
equal to half its radius of curvature. More precisely, we chose a
radius of curvature for M1 of 2000mm and a length of 50 mm, which
gives a waist for the pump cavity of 250 $\mu m$. The signal-idler
cavity had a M4 mirror of radius $R=100mm$. In the hemi-confocal
configuration, this gives a waist of 120 $\mu m$. The pump waist
is therefore twice as big as the signal-idler waist.

The injection system was also modified. The $TEM_{00}$ mode coming
out of the mode-cleaner was focussed into the OPO infrared cavity
with a waist size equal to three times the cavity waist. In the
transverse plane which is the reciprocal image of mirror M2
surface (which is the location of the minimum waist of the cavity
mode), we placed a black and white transmission pattern (USAF
resolution target, consisting of various horizontal or vertical
slit, numbers, squares). The field injected in the cavity is
therefore the image of the object chosen on the target multiplied
by a gaussian intensity envelope.

\subsection{Results}

The hemi-confocal cavity is not an exact self-imaging cavity
\cite{Arnaud}, meaning that all the transverse modes do not
resonate at the same length. Even without a nonlinear crystal
inserted in it, it is not a "transparent device", as it does not
perfectly transmit any input image. Its behavior in terms of image
transmission has been studied in detail in \cite{image}. It can be
described as follows : one finds four cavity resonances when
scanning the cavity length over a range $\lambda/2$. At each
resonance length, the cavity is resonant for one quarter of the
transverse modes. For example, if the cavity length is stabilized
on the resonance of the mode family containing the $TEM_{00}$
mode, it transmits the even part of the input image plus the even
part of its spatial Fourier transform.

\begin{figure}[htbp]
\begin{center}
\epsfig{file=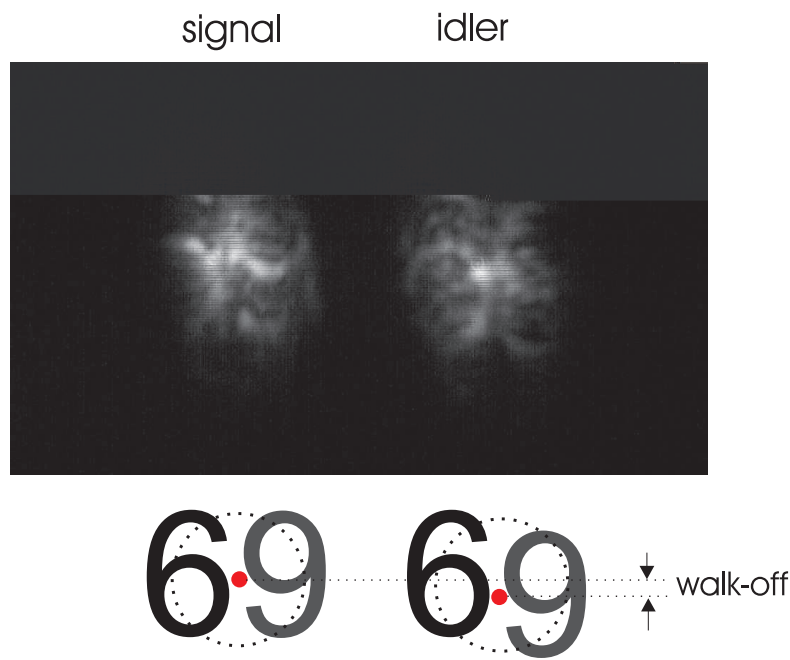,width=.8\linewidth} \caption{Above:
signal (left) and idler (right) images transmitted through a
hemi-confocal cavity without pump when an object consisting of an
off-axis "6" is injected in the cavity. Below: sketch of the
obtained pattern, indicating the optical symmetry axis (central
dots) and the waist size of the mode used to illuminate the object
(dashed circles), giving a kind of "field of view" in this
experiment, as the outer parts of the transmitted image cannot be
seen outside this circle} \label{walkoff}
\end{center}
\end{figure}

As an example, figure \ref{walkoff} shows the intensity patterns
recorded by the CCD camera on the signal and idler modes
transmitted through the "cold" hemi-confocal cavity (without pump
light). The input object is a "6" slightly shifted from the cavity
optical axis and injected with a polarization at 45 degrees from
the signal and idler polarizations. One observes at the output of
the cavity an image looking like a "69", multiplied by a gaussian
enveloppe limiting the field of view to the image center, plus a
bright pattern at the center: this complex feature corresponds
indeed to the even part of the input ("69") plus its spatial
Fourier transform (central part). Looking more precisely at the
signal and idler components of the output, one notices an effect
due to the walk-off: the transmitted image is slightly different
in the signal and idler beams. The "6" and the "9" are exactly
symmetric with respect to the central feature in the signal beam
and have a small lateral shift in the idler beam. This is due to
the fact that, because of the walk-off, the cavity axis is not
exactly the same inside the crystal for signal and idler.

\begin{figure}[htbp]
\begin{center}
\epsfig{file=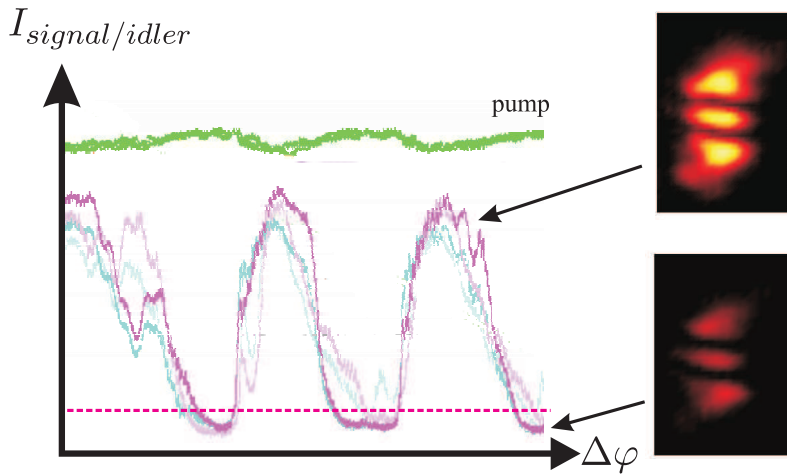,width=.8\linewidth} \caption{Left
side: pump, signal and idler total intensities as a function of
the relative phase between injected mode and pump mode. The dashed
line is an estimation of the signal and idler intensity without
amplification. Right side: transverse profile of the signal beam
at maximum and minimum amplification.}
 \label{amplidesampli}
\end{center}
\end{figure}

Taking a simpler image having the shape of a double slit
(multiplied by a Gaussian envelope) sent with a polarization at 45
degrees from the signal and idler polarizations, we have observed
what is, to the best of our knowledge, the first phase-sensitive
amplification of an image in the c.w. regime, as can be seen on
the total signal and idler intensities plotted on figure
\ref{amplidesampli} as a function of the relative phase between
the injected signal and the pump. From the constant value of the
transmitted pump and the almost equal values of the signal and
idler intensities, one notices that the triple resonance condition
is well satisfied during the whole recording. We have been able to
stabilize for the few seconds the relative phase on the maximum
and minimum amplification points using the transmitted
signal-idler modulation amplitude at 19.5MHz (due to the
corresponding pump modulation created by EOM2) as an error signal.
This has enabled us to record the whole transmitted image with the
CCD camera at the intensity maximum and minimum. These two images,
on the right side of figure (\ref{amplidesampli}), are both
roughly identical to the transmitted image without amplification,
that is not represented in the figure, but with different
intensities. Like in the example of figure (\ref{walkoff}) they
consist of the even part of the object (the slit double itself in
the present case) plus the vertical three dot pattern which is
central part of the Fourier transform of the horizontal double
slit. This last feature is here the most intense part of the image
because it is concentrated in the central part of the transverse
plane and is less affected by the overall multiplication by the
Gaussian beam shape.

\begin{figure}[tbp]
\begin{center}
\epsfig{file=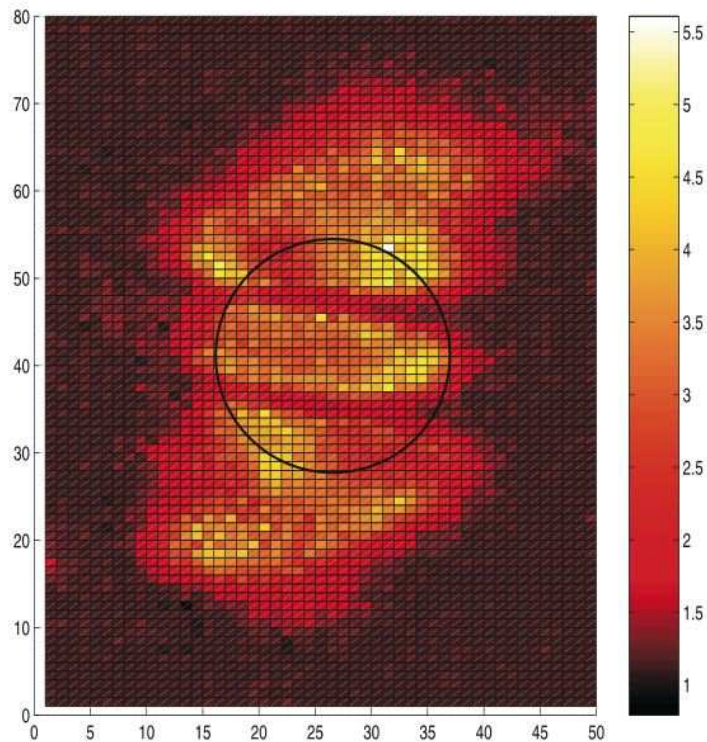,width=.8\linewidth}

\caption{Spatial variation of the ratio between the transmitted
images on the signal beam taken at the points corresponding to the
maximum and minimum total intensities. The circle indicates the
size of the $TEM_{00}$ cavity mode} \label{GainSpatial}
\end{center}
\end{figure}

By performing a point by point division of the two images, we
obtain an estimation of the local gain and of its spatial
variation. The result is represented on figure \ref{GainSpatial}.
For the same reasons as in the single mode case, it is not
possible to deduce from the data an exact value of the gain. One
notices that the calculated quantity is roughly homogeneous, with
a mean value close to 3 (5dB) over all the area where the local
intensity of the image has a significant value. The size of the
$TEM_{00}$ cavity mode is indicated by the black circle. The fact
that one gets significant amplification well beyond the extension
of this fondamental mode is another indication of the multimode
character of the present amplifier.

Finally we have found evidence of quantum correlations between the
signal and idler components of the amplified image. We have
observed that the noise level on the difference between the total
intensities of the signal and idler beams is 12 $\%$ below the
shot noise limit when the image mean intensity reaches its maximum
value. The reason why the quantum correlation turns out to be
smaller for the image amplifier than for the single mode amplifier
at comparable gains is unknown at the moment.

\section{Conclusion}

We have studied intracavity optical parametric amplification in
the c.w. regime, and obtained first results on a system which has
not received so far a great deal of attention, in spite of its
promising possibilities: evidence for single-mode phase
sensitive c.w. optical amplification with a significant gain in scanned cavity configuration,
observation of phase sensitive single mode amplification in the
quantum cloning regime, and more importantly first evidence of
c.w. phase sensitive amplification of simple images, also in the
quantum cloning regime.

These are encouraging first results which need to be improved and
investigated further. It would be in particular interesting to
determine the noise figure, and its spatial variation of this
amplifier. Let us note that the present set-up was not optimized
for single mode amplification, and that even better gain values
and more important quantum correlations could be obtained in a
more optimized configuration of finesses and cavity transmission
factors. Such an optimized system could also be used as a
noiseless amplifier.

As far as quantum imaging effects are concerned, the present
set-up does not enable us to look for local quantum correlations
and local squeezing because of the very low power of the
transmitted images, of the order of 100$\mu W$, so that
photocurrents measured on small portions of the image are buried
in the dark noise of the detection system. Phase sensitive
amplification in type-I configuration, which has already produced
impressive results in the single transverse mode case \cite{PKL} is
certainly a good candidate to produce measurable local quantum
effects in an image. This would be interesting, not only for
imaging, but also for their intrinsic quantum multimode
properties, since they open the way to multiplexing of quantum
information.

\section*{acknowledgements}

Laboratoire Kastler-Brossel, of the Ecole Normale Sup\'{e}rieure
and the Universit\'{e} Pierre et Marie Curie, is associated with
the Centre National de la Recherche Scientifique.\\
This work was supported by the European Commission in the frame of
the QUANTIM project (IST-2000-26019).


\begin{thebibliography}{99}

\bibitem{bachor}
N. Treps, U. Andersen, B. Buchler, P.K. Lam, A. Maître, H. Bachor,
C. Fabre, Phys. Rev. Letters \textbf{88}, 203601 (2002)

\bibitem{Jedrkiewisz}
O. Jedrkiewicz, Y.-K Jiang, E. Brambilla, A. Gatti, M. Bache, L. A. Lugiato, and P. Di Trapani
Phys. Rev. Lett. \textbf{93}, 243601 (2004)

\bibitem{Gatti:04}
 A. Gatti, E. Brambilla,
M. Bache, and L. A. Lugiato, Phys. Rev. A \textbf{70}, 013802
(2004)

\bibitem{heidmann}
 A. Heidmann, R. J. Horowicz, S. Reynaud, E. Giacobino, C. Fabre, and G. Camy, Phys. Rev. Lett. \textbf{59}, 2555 (1987)


\bibitem{Ducci:01}
S.Ducci, N. Treps, A. Maitre, and C. Fabre, Phys. Rev.
\textbf{64},023803  (2001)

\bibitem{Lugiato1}
   L. A. Lugiato, Ph. Grangier, JOSA B, Vol. \textbf{14} Issue 2 Page 225
(February 1997)

\bibitem{Petsas}
K.I. Petsas, A. Gatti, L.A. Lugiato, and C. Fabre,  Eur. Phys. J. D \textbf{22}, 501-512 (2003)


\bibitem{Martinelli}
M. Martinelli, N. Treps, S. Ducci, S. Gigan, A. Maitre and C.
Fabre, Phys. Rev. A, \textbf{67}, 023808 (2003)


\bibitem{Kumar:99}
 S.-K. Choi, M. Vasilyev, and P. Kumar,  Phys. Rev. Lett. \textbf{83},
1938 (1999)

\bibitem{Lantz}
E Lantz and F Devaux,
Quantum semiclass. Opt. \textbf{9}  279-286 (1997)

\bibitem{Levenson}
J.A Levenson, I. Abram, T. Rivera and P. Grangier, J. Opt. Soc. Am. B \textbf{10}, 2233 (1993)

\bibitem{Grangier}
 J. A. Levenson, I. Abram, T.
Rivera, P. Fayolle, J. C. Garreau, and P. Grangier,  Phys. Rev.
Lett. \textbf{70}, 267 (1993)

\bibitem{Protsenko} )I. Protsenko, L. Lugiato, C. Fabre, Phys. Rev \textbf{A50}, 1627
(1994)


\bibitem{Pfister} Sheng Feng, O. Pfister, Phys. Rev. Lett. \textbf{92}, 203601
(2004)

\bibitem{Laurat}
 J. Laurat, T. Coudreau, G. Keller, N. Treps, C. Fabre,quant-ph/0410081

\bibitem{Black:01}
Black E.D., Am. J. Phys. \textbf{69} 79 (2001)

\bibitem{Colville:94}
F. G. Colville , M. J. Padgett , and M. H. Dunn, Appl. Phys. Lett.
 \textbf{64} , 1490 (1994)

\bibitem{Wong:98}
J. Teja and Ngai C. Wong, Optics Express \textbf{2}, 65 (1998)

\bibitem{Rosencher:00}
B. Scherrer, I. Ribet, A. Godard, E. Rosencher, M. Lefebvre, JOSA
B, \textbf{17} 1716 (2000)

\bibitem{Turnbull:00}
G. A. Turnbull, D. McGloin, I. D. Lindsay, M. Ebrahimzadeh, M. H.
Dunn,  Optics Letters, Vol. \textbf{25} 341 (2000)

\bibitem{Arnaud}
J.A. Arnaud, Applied Optics,
Vol \textbf{8}. Issue 1, page 189 (1969)

\bibitem{image}
S. Gigan, L. Lopez, N.Treps, A. Maître, C. Fabre, Image transmission through a stable paraxial cavity, Preprint.

\bibitem{PKL}
 T. C Ralph and P. K. Lam, Phys. Rev. Lett. 81, 5668 (1998)

\end{thebibliography}
\end{document}